# A Mathematical Model to Elucidate Brain Tumor Abrogation by Immunotherapy with T11 Target Structure

Sandip Banerjee[1]☯*, Subhas Khajanchi[1]☯, Swapna Chaudhuri[2]

**1** Department of Mathematics, Indian Institute of Technology Roorkee, Roorkee - 247667, Uttaranchal, India,
**2** Department of Laboratory Medicine, School of Tropical Medicine, Kolkata-700073, West Bengal, India

☯ These authors contributed equally to this work.
* sandofma@iitr.ac.in

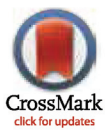

## Abstract

T11 Target structure (T11TS), a membrane glycoprotein isolated from sheep erythrocytes, reverses the immune suppressed state of brain tumor induced animals by boosting the functional status of the immune cells. This study aims at aiding in the design of more efficacious brain tumor therapies with T11 target structure. We propose a mathematical model for brain tumor (glioma) and the immune system interactions, which aims in designing efficacious brain tumor therapy. The model encompasses considerations of the interactive dynamics of glioma cells, macrophages, cytotoxic T-lymphocytes (CD8+ T-cells), TGF-$\beta$, IFN-$\gamma$ and the T11TS. The system undergoes sensitivity analysis, that determines which state variables are sensitive to the given parameters and the parameters are estimated from the published data. Computer simulations were used for model verification and validation, which highlight the importance of T11 target structure in brain tumor therapy.

## Introduction

In the rank of deadly form of tumors, adult primary malignant gliomas are the most common primary brain tumors, occurring at a rate of five cases per 1,00,000 population per year [1]. The survival rate for malignant gliomas in the category of Grade-IV and Grade-III varies from 1 year to 3 years respectively [2, 3]. These high grade gliomas are not differentiable and are genomically unstable. They have infiltrative behavior in their sequestered location beyond the blood-brain-barrier (BBB). Therefore, conventional treatments including surgery, radiation and chemotherapy often fails to control malignant gliomas, namely, Glioblastoma Multiforme (GBM), which is the most common malignant glioma. Hence, there is a need for novel therapies, namely, immunotherapy, in the hope that there is an increase in the survival rate of the patients.

Using mathematical modeling as a viable tool, complex biological processes are studied. Because of the complexity and unpredictable pattern of the gliomas, mathematical modeling can be extremely helpful in analyzing factors that may contribute to the complexity intrinsic in insufficiently understood glioma development process. Researchers have developed several types of GBM models in recent years. The theoretical study of gliomas supported by experimental finding has been classified mainly into two categories. While one group of researchers study the temporal and spatiotemporal dynamics of glioma proliferation and invasion, the other group formulates new therapies as a treatment, that may result in the survival of patients with high grade gliomas.

Many mathematical models that describe the temporal or spatiotemporal dynamics of glioma proliferation and invasion have been formulated. Through mathematical modeling, it is possible to answer the diverse biological questions concerning the analysis of early GBM growth, therapy effectiveness or even simulations in realistic brain structure. The mathematical model developed by Swanson et al. quantifies the spatio-temporal proliferation and invasion dynamics of gliomas in a 3D diffusion framework. Their model portrays the growth and extension of theoretical glioblastoma cells in a matrix that accurately describes the brain's anatomy to a resolution of 1 cubic millimeter. The model, not only has a significant resemblance with the magnetic resonance imaging (MRI) of actual patients but also shows the distribution of diffusely infiltrating cells [4–8]. Eikenberry et al., in their work, predicted patterns of tumor recurrence following various modes of therapeutic intervention through three-dimensional mathematical model [9]. The first 3D model of solid glioma tumor growth, by developing a cellular automata was done by Kansal et al., which realistically models the macroscopic behavior of a malignant tumor using predominantly microscopic parameters [10]. Agent based modeling was also used to realistically simulate early GBM growth. The simulation provides insight into the invasive nature of the GBM, its average invasion speed that drives the tumor to spatial expansion [11]. Toma et al. [12] modeled brain tumor growth at the cellular level considering the effect of microglial cells on the progression of malignant primary brain tumor with the help of partial differential equations. The qualitative results presented in their work are in agreement with *in-vitro* data.

One of the dominant aspects of malignant glioma growth is the invasion of brain, which prioritizes the modeling of invasion dynamics. A theoretical framework of invasion of brain tumor was introduced first by Tracqui et al. [13], followed by Woodward et al. [14] and Burgess et al. [15]. Glioma invasion into a non-homogeneous brain structure was also studied by Swanson et al. [4] using the BrainWeb brain atlas. The authors simulate a realistic brain geometry including fibre differentiation into grey and white matter and compare macroscopic simulation data with clinical data obtained from the analysis of a series of one patient's CT scans. Wurzel et al. made a different approach and developed a cellular automation model, that simulates the invasion, proliferation and death of tumor cells [16]. Kim et al. [17] presented a mathematical model of glioblastoma multiforme evolution, its relative balance of growth and invasion. Their model succeeded in providing explanation for the growth/invasion cycling patterns of glioma cells in response to high/low glucose uptake in micro-environment *in-vitro* and suggests new target for drugs, associated with miR451 upregulation.

Many GBM patients are subjected to glioma chemotherapy, tumor resection and radiation therapy [18, 19] at some point during the course of their disease. The first modeling attempt of glioma chemotherapy was done by Tracqui et al. [13], where the authors considered a patient suffering from anaplastic astrocytoma treated with chemotherapy and modeled chemotherapy as a spatially homogeneous equation of a reaction diffusion system. The model simulation agreed with the clinical data derived from CT scans. Swanson et al. [6] proposed a modified model for chemotherapy of GBM tumors, which assumes brain structure to be heterogeneous (grey and white matter differentiation) that makes diffusion process space dependent. The simulations which were performed for realistic brain geometries, showed the decay of tumor cells in grey matter due to sufficient drug delivery but failed to control the proliferation of glioma cells in the white matter. This model incorporates all the properties of the Tracqui model and gave better understanding of the therapy effect.

Modeling the effects of various types of tumor resection were studied by many groups. Woodward et al. [14] made the most noticeable study on resection modeling, where they used the methodology of Tracqui et al. [13] (reaction diffusion equation). The system parameters were estimated on the same patient's CT scan of Tracqui et al. [13]. Several scenarios of GBM resection were investigated by the authors and simulation of the model showed tumor recurrence. Swanson et al. [7] extended the work of Woodward et al. by considering inhomogeneous brain structure. Numerical simulation of the model with different diffusion and growth rates of the malignant glioma cells showed consistent recurrence of the resected tumors.

Many types of cancer, including brain tumor, have been subjected to systemic immunotherapy by exogenous administration of immune cells or immunoregulatory factors with limited success. Chakrabarty and Hanson formulated an optimal control problem of drug delivery to brain tumor to minimize the tumor cell density and reducing the side effect of drugs using Galerkin finite element method [20]. Bandara et al. proposed a mathematical model to support the rational development of targeting strategies (in-silico) for efficiently transporting Paclitaxel, an antimitotic drug across the blood-brain-barrier [21]. Kirkby et al. formulated a model of glioblastoma, which predicts the effects of escalator of radiotherapy dose and reproduces extremely accurate clinical data [22]. Schmitz et al., through a cellular automaton model of brain tumor studied heterogeneous tumors with both treatment sensitive and treatment resistant cells. The authors investigated the monoclonal tumors, two-stream with resistant subpopulation and multi-strain tumors with induced mutation and able to present survival time data from each of these case studies [23]. Walker and Cook [24], through macroscopic model, developed a system of drug delivery to brain tumors, where the authors have assumed that the drug is delivered to both normal and tumor tissue through vasculature system. By means of diffusion and convection, the drugs penetrate the brain tissue across the vasculature and transported across the blood-brain-barrier and through the interstitial space. By assessing the effects of changing parameters on drug delivery, they found an optimal treatment for convective drug delivery to the center of the tumor. Kronik et al. [25] considered the interactive dynamics of cytotoxic T-Lymphocytes, brain tumor cells, major histocompability complex (MHC) class-I and MHC class-II along with cytokines TGF-$\beta$ and IFN-$\gamma$, where they have used computer simulation for model verification and retrieving putative treatment scenarios. Their result suggests that GBM may be eradicated by new dose-intensive strategies or by significantly encouraging the endogenous immune response or by T-cell infusion, as shown by their mathematical model.

In this paper, we present a mathematical model for immunotherapy of brain tumor with T11 target structure. The model describes a complex interaction involving glioma cells, macrophages, cytotoxic T-lymphocytes, the immuno-suppressive component TGF-$\beta$ and the cytokine IFN-$\gamma$. We use published data [26] to analyze the sensitivity of the system parameters and to estimate their values. The model is then used to validate various immunotherapeutic scenarios as observed in [26].

## Materials and Methods

### Biological Framework and Experimental Facts

The neurotoxic role of N-ethyl-N-nitrosourea (ENU) in inducing CNS (central nervous system) tumors is well-established, but the assessment of a gradual transformation of the normal cells to cancer cells has not been studied. In this experiment [26], the study elucidates the changes on rate of survival, growth kinetics, immunological and histological parameters chronologically from ($2^{nd}$–$10^{th}$) months after induction of glioma in rats from neonatal stage. Healthy newborn Druckray rats of both sexes were maintained in our institutional animal facility. Six animals in each group were weaned at 30 days of age and housed in separate cages at 22°C in a 12 h light/darkness cycle. Animals were fed autoclaved rat feed pellets and water ad libitum. The experimental animals were grouped into five groups. (i) N: age-matched normal healthy control. (ii) ENU: 3–5-days-old neonatal animals injected with ENU intraperitoneally (i.p.) (iii) ET1: 5-month-old ENU-treated animals injected(i.p.) with the first dose of T11TS. (iv) ET2: 5-month-old ENU-treated animals injected (i.p.) with the first and second doses of T11TS at a 6-days interval. (v) ET3: 5-month-old ENU-treated animals injected (i.p.) with the first, second, and third doses of T11TS at 6-days interval between each dose. Maintenance and animal experiment procedures were strictly done as per the guidelines of the Committee for the Purpose of Control and Supervision of Experiments on Animals (CPCSEA), Government of India and was monitored by the Institutional Animal Ethics Committee (IAEC), Post Graduate Education and Research, Kolkata, India. Care was taken to minimize animal usage and not inflicting pain to the animals in study. The control and experimental rats were euthanized by over dose of Ketamine Hcl (80 mg/kg body wt, route: intramuscular). The Institutional Animal Ethics Committee (IAEC), Post Graduate Education and Research, Kolkata, India, specifically approved this study. Furthermore, an attempt was made to investigate the immunomodulatory and antitumor properties of sheep red blood cells (SRBCs), glycopeptide known as T11 target structure (T11TS).

3–5 days old Druckray rats of both sexes were glioma induced by intraperitoneal (i.p.) injection of ENU. $2^{nd}$, $4^{th}$, $6^{th}$, $8^{th}$ and $10^{th}$ months of rats after brain tumor (glioma) induction were sacrificed to study tumor growth kinetics and a gamut of immunological parameters like rosette technique of lymphocytes and phagocytosis peripheral macrophages and polymorphonuclear neutrophil (PMN). Survival studies were done from the different groups of animals. The survival study after ENU was injected included observations, which was done by registering the total number of days an individual animal survived and the mean survival time in each group. The survival rate of animals were also recorded after $2^{nd}$, $4^{th}$, $6^{th}$, $8^{th}$ and $10^{th}$ month of ENU induction.

Proliferation index and the fluorochrome uptake studies were performed at an interval of $2^{nd}$ month up to $10^{th}$ month in ENU induced animals. Portions of tumor susceptible areas from all the groups of rats were taken as samples of primary explant technique from which serial passages were made to obtain a steady culture from which the experiments were performed. Splenic macrophages were separated by adherent technique and polymorphonuclear cells (PMN) separated by density gradient with percoll were subjected to nitroblue tetrazolium assay to determine the phagocytic capacity of the individual cells in different groups. For evaluating ENU induced cytotoxicity, cytolytic-asssays of splenic lymphocytes (SL) were performed for HO-33342 release in the different groups of rats. Tumor cells (a steady glioma tumor line, syngenic in nature) were tagged with HO-33342 fluorochrome dye (total incorporation) for 15 minutes at 37°C and the excess untagged dye was washed off. Fluorochrome released due to target lysis was measured in a spectrofluorimeter.

Immunological findings have been strongly correlated with the histological findings. Administration of the T11TS in ENU treated animals demonstrated the anti-neoplastic activity in a dose dependent manner, as evidenced through the conversion of neoplastic glial cytopathology to normal glial features. After 6 months of ENU induction, Grade-IV oligodendroglioma was observed with mitotic figure, giant cells and a minimum intercellular space. Following the $10^{th}$ month after ENU administration, the histological observation of brain sections showed a ribbon like appearance of closely packed dividing cells, degenerative fibrils, oligodendroglioma Grade-IV, mixed glioma. The $1^{st}$ dose of the T11TS fraction shows reduced glioma cells due to apoptotic death. Reversion of neoplastic glioma features to normal glioma features was observed following the 3<sup>rd</sup> dose of the T11TS fraction.

## Model Formulation

The aim of the mathematical model is to yield a simplified version of the complicated biological processes. The analytical power of the model can be greatly enhanced by deliberately isolating the crucial forces in the system and neglecting the secondary effects. Our mathematical model describes the effect of interactions between glioma cells (brain tumor cells) and the immune system, which includes macrophages, $CD8^+$ T-cells, TGF-$\beta$ and IFN-$\gamma$. The model focusses on the role of T11 target structure (T11TS) along with immune system as immunotherapy to brain tumor, with the aim to simulate and thus evaluate possible therapeutic scenarios.

As the brain tumor (glioma) grows in size, it starts secreting immune suppressive factors like Transforming Growth Factor-$\beta$ (TGF-$\beta$), Prostaglandin E2(PG-E2), Interleukin-10(IL-10), etc. These factors cross the blood-brain-barrier (BBB) and reach the peripheral immunocytes, making them functionally inactive, so there is an overall immune suppressive state, helping in the growth of the tumor. The tumor antigens also cross the BBB making the lymphocytes specific for the tumor antigens, that is, they will attack the glioma cells once they encounter. At this juncture, when T11 Target Structure (T11TS) is injected intra-peritoneally, it activates the systemic lymphocytes and also the macrophages. The activated macrophages cross the BBB and enter the brain to attack the glioma cells and kill them by phagocytosis. The lymphocytes once activated by T11TS enter the nearest lymph-node and crosstalk with macrophages which present them as tumor antigens, making them specific for the killing of glioma cells. Both CD4 + and CD8+ T-lymphocytes now can cross the large blood barrier at an increased number. They now crosstalk with the macrophages within the brain, which again present them as a tumor antigens. The macrophages and the lymphocytes after their crosstalk secrete various cytokines which are conducive for the elimination of glioma cells. The macrophages and the lymphocytes, which are now activated and armed for glioma killing, eliminates glioma cells by apoptosis. A schematic diagram depicting the above mentioned biological scenario is shown in [Fig 1](#).

In our model we consider a simplified version of the schematic diagram. Our goal is to formulate a mathematical model which allows sufficient complexity so that the model may qualitatively generate the tumor growth pattern, while it simultaneously maintains sufficient simplicity for analysis. The model we propose is a system of five non-linear ordinary differential equations (ODEs), to characterize the dynamics of the interaction between the glioma cells (G) and different immunological components, namely, macrophages (M), cytotoxic T-lymphocytes or CD8+T cells ($C_T$), TGF-$\beta$ ($T_\beta$), IFN-$\gamma$ ($I_\gamma$) and the external anti-tumor agent T11TS ($T_s$).

**Dynamics of Glioma Cells.** The dynamics of glioma cells is given by

$$\frac{dG}{dt} = r_1 G \left(1 - \frac{G}{G_{max}}\right) - \left(\frac{1}{T_\beta + e_1}\right)(\alpha_1 M + \alpha_2 C_T)\left(\frac{G}{G + k_1}\right) \qquad (1)$$

We assume that in the absence of immune system glioma cells follow logistic growth, given by the first term of the [Eq (1)](#); $r_1$ is the intrinsic growth rate of glioma cells and $G_{max}$ is its carrying capacity, that is, the maximal tumor cell burden. The second term of the [Eq (1)](#) shows how glioma cells are eradicated by the macrophages and CD8+T cells at the rates $\alpha_1$ and $\alpha_2$ respectively. It is also assumed that the elimination of the glioma cells by macrophages and CD8+T cells are proportional to both G, M and G, $C_T$ respectively, with saturation for large G. Michaelis-Menten term is being incorporated to bring out the accessibility of the glioma cells to macrophages and CD8+T cells, implying that the effect of macrophages and CD8+T cells efficacy follow Michaelis-Menten saturation dynamics, $k_1$ being the half saturation constant. The term $\frac{1}{T_\beta + e_1}$ is the major immunosuppressive factor for the activity of both the macrophages and CD8+T cells, $e_1$ being the Michaelis constant.

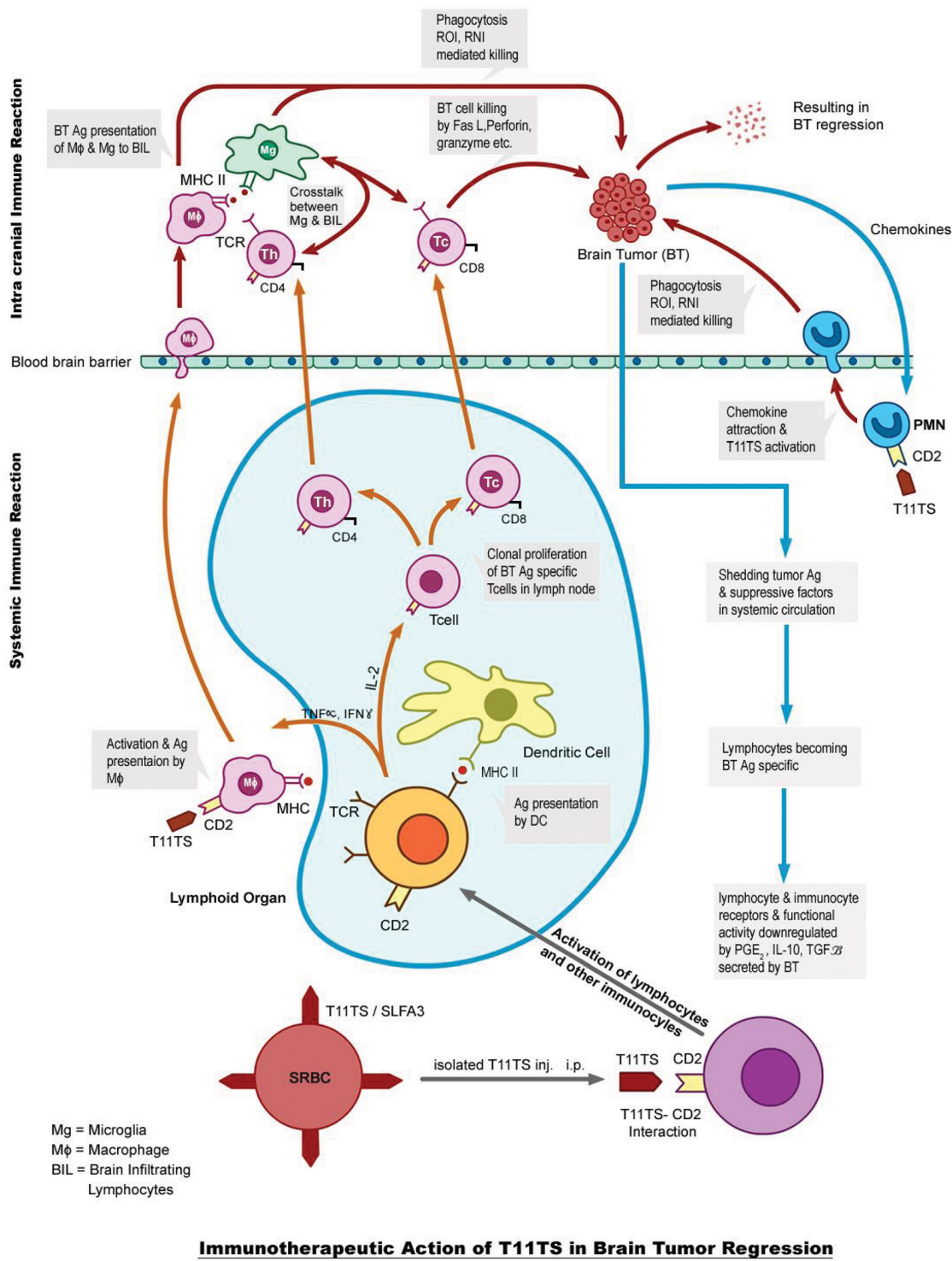

**Fig 1. Schematic diagram.** The figure shows the dynamics between brain tumor and the immune components, namely, macrophage, microglia, CD4+ T cells, CD8+ T cells, dendritic cells, TGF-$\beta$, IFN-$\gamma$ and the immunotherapeutic agent T11 target structure.

**Dynamics of Macrophages.** Dynamics of macrophages (M) is described by the Eq (2):

$$\frac{dM}{dt} = r_2 M \left(1 - \frac{M}{M_{max}}\right) + a_1 \left(\frac{I_\gamma}{k_4 + I_\gamma}\right) \left(\frac{1}{T_\beta + e_2}\right) - \alpha_3 \left(\frac{G}{G + k_2}\right) M \quad (2)$$

Since, the macrophage growth pattern is not evident from existing literatures, we assume that the macrophages grow logistically (in absence of glioma cells) with intrinsic growth rate $r_2$ and carrying capacity $M_{max}$ (given by the first term of the Eq (2)). The second term in Eq (2) shows the activation of macrophages by IFN-$\gamma$ at a rate $a_1$, $k_4$ being the half saturation constant, implying the presence of Michaelis-Menten saturation dynamics. At the same time, the term $\frac{1}{T_\beta + e_2}$ interrupts the activity of macrophages, it is a degradation term with saturation constant $e_2$. The third term gives the rate of immuno-induced macrophage death by malignant glioma cells at the rate $\alpha_3$, $k_2$ being the half saturation constant standing for the accessibility of glioma cells to macropahges.

**Dynamics of CD8+ T Cells.** CD8+ T cell dynamics is given by

$$\frac{dC_T}{dt} = \frac{a_2 G}{k_5 + T_\beta} - \mu_1 C_T - \alpha_4 \frac{G}{G + k_3} C_T \quad (3)$$

The first term in the Eq (3) represents the recruitment term of CD8+T cells which occurs due to the direct presence of malignant glioma cells, $a_2$ being the antigenicity of the glioma cells which triggers an immune response in the host. The immunosuppressive response by TGF-$\beta$ puts limitation to the recruitment level, $k_5$ being termed as the inhibitory parameter. $\mu_1$ is the rate of loss of CD8+T cells due to inflammatory reaction in the brain. The last term in Eq (3) represents the clearance of CD8+T cells by the glioma cells at a rate $\alpha_4$, $k_3$ being the half saturation constant.

**Dynamics of the Cytokine TGF-$\beta$.** Experimental evidence [27] suggests that TGF-$\beta$ is produced in a small quantity when production of glioma cells is small but it gets ample nutrient from the neighboring tissue. But when glioma cell production grows sufficiently large resulting in lack of oxygen and space, it starts producing TGF-$\beta$ to stimulate angiogenesis and to destroy immune response for tumor growth [28]. Eq (4) describes the dynamics of TGF-$\beta$:

$$\frac{dT_\beta}{dt} = s_1 + b_1 G - \mu_2 T_\beta \quad (4)$$

The first term in Eq (4) represents the constant source term of TGF-$\beta$ and the second term is the source term which is proportional to the glioma size, $b_1$ being the release rate per glioma cells. The last term is the degradation of TGF-$\beta$ at a constant rate $\mu_2$.

**Dynamics of the Cytokine IFN-$\gamma$.** Eq (5) represents the dynamics of IFN-$\gamma$:

$$\frac{dI_\gamma}{dt} = b_2 C_T - \mu_3 I_\gamma \quad (5)$$

IFN-$\gamma$ activates the macrophages, which are capable of destroying the glioma cells. We assume that CD8+T cells is a source of IFN-$\gamma$ [29] given by the first term of the Eq (5). The second term shows the degradation of IFN-$\gamma$ at a constant rate $\mu_3$.

After the administration of immunotherapeutic agent T11TS, which activates different immunocytes including macrophages, CD8+ T cells, we assume that these doses of T11TS will have an effect on the cell count of macrophages and CD8+ T cells [26]. In mathematical terms, we represent the administration of doses of T11TS by Dirac Delta function as input $\delta(t - \tau)$, where $\delta(t - \tau) = 1$, $t = \tau$ and 0, elsewhere. Three doses of T11TS were administered in the system, the first one in the $7^{th}$ month ($210^{th}$ day $= \tau_1$), followed by an interval of 6 days [26], that is, in the $216^{th}$ day ($\tau_2$) and $222^{nd}$ day ($\tau_3$).

The interactions of immune system, the dynamics of glioma cells and T11TS bringing together, leads to the following coupled non-linear ODEs:

$$\frac{dG}{dt} = r_1 G \left(1 - \frac{G}{G_{max}}\right) - \left(\frac{1}{T_\beta + e_1}\right)(\alpha_1 M + \alpha_2 C_T)\left(\frac{G}{G + k_1}\right)$$

$$\frac{dM}{dt} = r_2 M \left(1 - \frac{M}{M_{max}}\right) + a_1 \left(\frac{I_\gamma}{k_4 + I_\gamma}\right)\left(\frac{1}{T_\beta + e_2}\right) - \alpha_3 \left(\frac{G}{G + k_2}\right) M + T_s \sum_{i=1}^{3} \delta(t - \tau_i)$$

$$\frac{dC_T}{dt} = \frac{a_2 G}{k_5 + T_\beta} - \mu_1 C_T - \alpha_4 \frac{G}{G + k_3} C_T + T_s \sum_{i=1}^{3} \delta(t - \tau_i) \qquad (6)$$

$$\frac{dT_\beta}{dt} = s_1 + b_1 G - \mu_2 T_\beta$$

$$\frac{dI_\gamma}{dt} = b_2 C_T - \mu_3 I_\gamma$$

with initial conditions

$$G(0) = G_0 \geq 0, \quad M(0) = M_0 \geq 0, \quad C_T(0) = C_{T0} \geq 0, \quad T_\beta(0) = T_{\beta0} \geq 0, \quad I_\gamma(0) = I_{\gamma0} \geq 0.$$

## Approaches to the estimation of system parameters

Estimating parameter values for the constructed mathematical model is one of the hardest challenge which a modeler faces. Various techniques are approached to simplify the process. Sensitivity analysis is one such tool which simplifies the process of parameter estimation into several stages that are easier to solve. By using the sensitivity analyzing technique, the parameters which are less sensitive as well as unidentifiable are categorized. The set of highly sensitive parameters, which has been reduced to a tractable size, can be estimated from the available experimental data.

### The sensitivity analysis

In our model, we have 23 parameters, for which we need to determine its sensitivity and rank the parameters with respect to their identifiability. The sensitivity graph obtained using the code **myAD**, developed by Prof. Martin Fink [30], is shown in the Fig 2. To quantify the sensitivity from the figure, we calculate the sensitivity coefficient by non-dimensionalizing the sensitivity functions and computing $L^2$ norm of the resulting functions, given by

$$C_{ij} = \left\|\frac{\partial u_i}{\partial q_j}\frac{q_j}{max(u_i)}\right\|_2^2 = \int_{t_0}^{t_f}\left|\frac{\partial u_i}{\partial q_j}\frac{q_j}{max(u_i)}\right|^2 dt \qquad (7)$$

After comparing and ranking the sensitivity function we can sort the most sensitive parameters (in descending order) to the least ones as shown in the bottom panel of Fig 2. We next define Fisher's information matrix $F = S^T S$. We compute the normalized sensitivity function matrix S

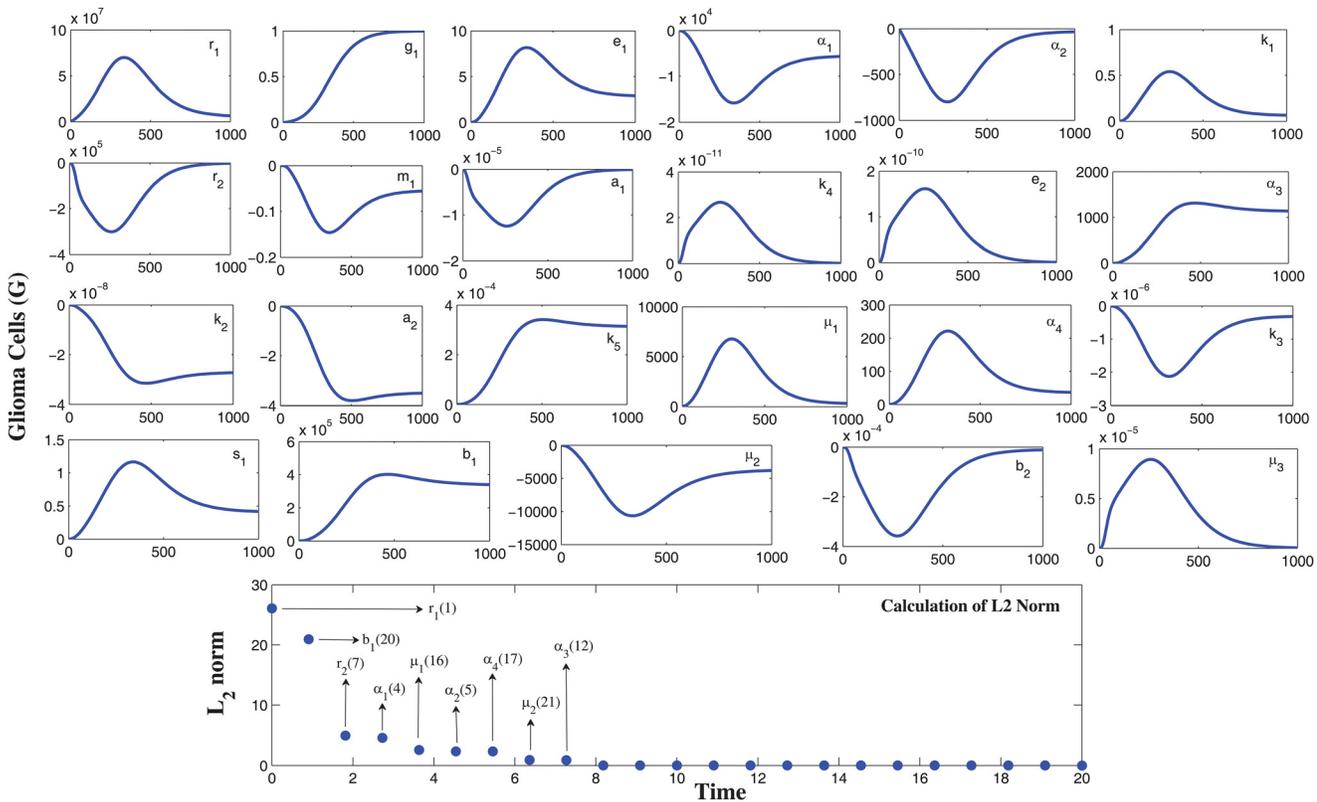

**Fig 2. Relative sensitivities of the parameters using automatic differentiation.** From the top panel figures, it is observed that the parameters $r_1$, $\alpha_1$, $\alpha_2$, $r_2$, $\alpha_3$, $\mu_1$, $\alpha_4$, $b_1$, $\mu_2$ are sensitive with respect to the malignant glioma cells. The observation window is [0, 1000] and the sensitivity of a parameter is identified by the maximum deviation of the state variable (along y-axis) and it also identifies the time intervals when the system is most sensitive to such changes. The bottom panel gives the sensitivity quantification by calculating sensitivity coefficient through $L^2$ norm.

doi:10.1371/journal.pone.0123611.g002

using automatic differentiation [30]. The numerical rank of $S^T S$ was computed to be 9 (taking square root of machine precision epsilon $\epsilon = 10^{-8}$). We next use the QR factorization technique with the column pivoting which is implemented in the MATLAB routine **qr**, [Q,R,P] = **qr**(F). This method determines a permutation matrix $P$ such that $FP = QR$ (QR being the factorization of $FP$). The indices in the first $k$ columns of $P$, identify the $k$ parameters that are most estimable. Our case yields the ordering $[r_1, b_1, r_2, \alpha_1, \mu_1, \alpha_2, \alpha_4, \mu_2, \alpha_3]^T$, that is, out of the 23 parameters, only the first nine ranked parameters are the most identifiable and sensitive parameters. Please note that, we will also estimate the parameter values $a_1$ and $k_5$ from the available data as we are unable to get them from any other sources. Since, they are not sensitive, their estimated values will not affect the dynamics of the system much.

## The estimation of parameters

The analysis and behavior of a mathematical model, to describe a given system depends on the system parameters. After identifying the most sensitive parameters, we now estimate the value of the system parameters in the following manner:

TGF-$\beta$ has a hepatic half life of 2.2 minutes [31]. However, the actual brain TGF-$\beta$ will have a slower decay rate because of its distance from the liver and its necessity to pass through the blood-brain-barrier (BBB). This time was estimated to be 6 minutes [4], which is in hours $\approx$ 0.1 hours. Therefore, the decay rate is given by, $\mu_2 = \frac{\log 2}{0.1h} = 6.931431 h^{-1} \approx 6.93 \ h^{-1}$.

In the Cerebral Spinal Fluid (CSF) of a glioblastoma patient, Peterson et al. [32] found the concentration of TGF-$\beta$ to be 609 pentagram (pg)/ml. We assumed the volume of CSF to be 150 ml. Since, there is a production of the tumor in a healthy individual, we obtain at steady state, $s_1 - \mu_2 T_\beta = 0$, which implies $s_1 = \mu_2 T_\beta = 6.93 h^{-1} \times 60.9 \ \frac{pg}{ml} \times 150 \ ml = 63305.55 \ pg \ h^{-1} \approx 6.3305 \times 10^4 \ pg \ h^{-1}$ (in a healthy person the concentration of TGF-$\beta$ is 10 times less than a glioblastoma patient).

For patients suffering from GBM, the mean level of cytokine TGF-$\beta$ is 609 pg(ml)$^{-1}$ = 609 pg(ml)$^{-1} \times 150$ ml = 91350 pg [32]. Using the equation at the steady state, we have,

$$b_1 = \frac{\mu_2 T_\beta - s_1}{G} = \frac{6.93 \ h^{-1} \times 91350 \ pg - 63305.55 \ pg \ h^{-1}}{10^{11}}$$
$$= 0.000005697 \ pg \ h^{-1} cell^{-1} \approx 5.70 \times 10^{-6} cell^{-1} pg \ h^{-1}.$$

The median half life of IFN-$\gamma$ is found to be 0.283 days = $(24 \times 0.283)$ h = 6.792 h $\approx$ 6.8 h [33]. Therefore, the hourly degradation of IFN-$\gamma$ is given by

$$\mu_3 = \frac{\log_e 2}{6.8 \ h} = \frac{0.693}{6.8 \ h} = 0.101933409 h^{-1} \approx 0.102 h^{-1}.$$

The production rate of IFN-$\gamma$ by a single $CD8^+T$ cell, $b_2$ is obtained (at the steady states) by

$$b_2 = \frac{\mu_3 I_\gamma}{C_T} = \frac{0.102 h^{-1} \times 200 \ pg/ml}{2 \times 10^5 \ cells/ml} = 0.000102 \ pg \ cells^{-1} h^{-1} = 1.02 \times 10^{-4} \ pg \ cells^{-1} h^{-1}$$

where 200 pg ml$^{-1}$ of IFN-$\gamma$ is reported in Kim et al. [34] and we assume $2 \times 10^5$ CTL ml$^{-1}$.

The range for antigenicity of glioma cell ($a_2$) is taken to be (0–0.5) obtained from Rosenberg et al [35]. The half life of CD8+T cells is reported to be 3.9 days [36], therefore the estimated hourly death rate $\mu_1$ is given by

$$\mu_1 = \frac{\log_e 2}{(3.9 \times 24) \ h} \approx 0.0074 \ h^{-1}.$$

Activated CD8+T cell eradicates (0.7 -3) target cells per day [37]. Assuming that a CD8+T cells kills 2.5 target cells per day, the rate is calculated to be $\frac{2.5}{24} \approx 0.1042$ cells h$^{-1}$. This experiment was done with $5 \times 10^5$ glioma cells/ml in a separated culture disc [26]. The rate at which activated CD8+T cells kill the glioma cells is estimated to be 0.1739 h$^{-1}$ [26]. Therefore, we get

$$\alpha_4 \frac{G}{G + k_3} = 0.1042 \ h^{-1} \Rightarrow k_3 = 3.34452 \times 10^5 cells.$$

From Mukherjee et al. [26] (namely, Fig 3, page 329), we obtain the cell proliferation index of glioma cells, before and after the administration of T11 target structure. It is observed that there is a steep increase of cell proliferation index when 2$^{nd}$, 4$^{th}$, 6$^{th}$ months old ENU injected animal brain cells, whereas in the 8$^{th}$ and 10$^{th}$ months, the increase is not very significant compared to previous months. We obtain the cell-count, starting from 0$^{th}$ month till the 10$^{th}$ month (6 data-points) and these, we consider to be the growth of glioma cells in the absence of the immune system. The parameters to be estimated with these 6 data-points are $r_1$ and $G_{max}$. To start the estimation process, the initial values of the parameters are chosen arbitrary (within meaningful biological range). We use the least square method to minimize the sum of the residual, to obtain the estimated values of the system parameters $r_1$, $G_{max}$. In practice, we use the

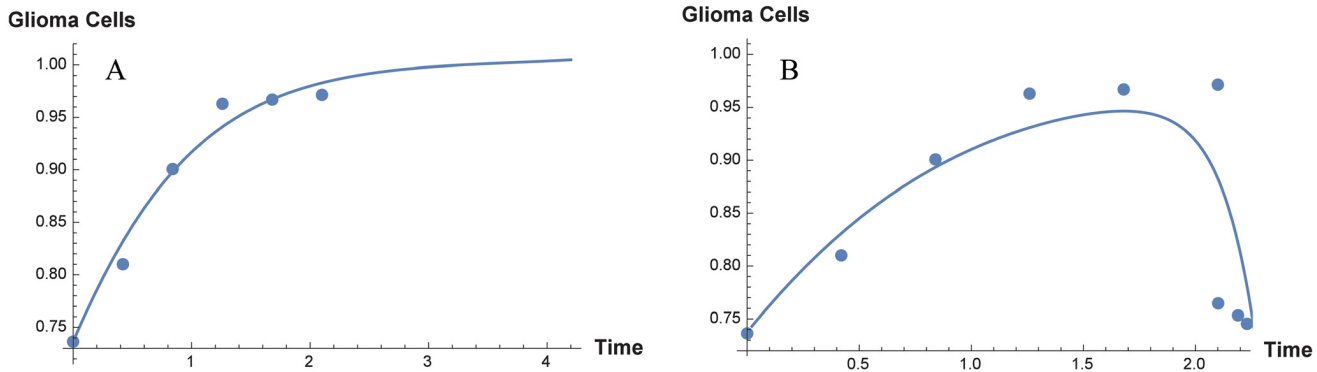

**Fig 3. System parameter estimation by the method of least squares.** Six data points are used to estimate the parameters $r_1$ (glioma growth rate) and $G_{max}$ (carrying capacity) in absence of immune system. The left panel A shows the best fit curve for the estimation of the parameters $r_1$ and $G_{max}$. The parameters $r_2$, $a_2$, $\alpha_1$, $\alpha_3$, $\alpha_4$ and $k_5$ are estimated by using nine data points obtained during immunotherapy by T11TS, the right panel B showing the curve of best fit.

doi:10.1371/journal.pone.0123611.g003

MATLAB function fminsearch to estimate the parameter values. The estimated values of $r_1$ and $G_{max}$ are found to be respectively, 0.01 h$^{-1}$ and $8.8265 \times 10^5$ cells. Fig 3A shows the best fit estimate for the model parameters $r_1$ and $G_{max}$. Next we add 3 more data points from the glioma cell proliferation index obtained after receiving 3 doses of T11TS (see Fig 3, page 329 of [26]), which shows a significant decrease in the cell count. We now consider the overall 9-data points reflecting the dynamics of glioma cells and the immune system (including the immunotherapy of T11TS). The same process is repeated and we estimate the following parameters: $r_2$ = 0.3307, $a_1$ = 0.1163, $\alpha_1$ = 1.5, $\alpha_3$ = 0.0194, $\alpha_4$ = 0.1694, $k_5 = 2 \times 10^3$. Fig 3B shows the best fit estimate for the model parameters $r_2$, $a_1$, $\alpha_1$, $\alpha_3$, $\alpha_4$ and $k_5$. Please note that the whole estimation process using the method of least squares is done by non-dimensionalizing the proposed model. The non-dimensionalized values obtained are then converted back to the actual parameters.

We have estimated the intrinsic growth rate ($r_1$), carrying capacity of glioma cells ($G_{max}$), killing rate of macrophage ($\alpha_1$) as 0.01 h$^{-1}$, $8.8265 \times 10^5$ cells, 1.5 h$^{-1}$ respectively, by fitting model predictions from the result of Mukherjee et al. [26]. We took the value of $e_1$ from Peterson et al. [32], which is the order of magnitude of the base line. The term is multiplied by the volume of CNS, given by $e_1$ = 150 ml × 60.9 pg/ml = 9135 pg $\approx 10^4$ pg. The value of $k_1$ (half saturation constant) is $2.7 \times 10^4$ cells [26].

Activated CD8+T cell kills the target cells (0.7–3) per day, reported by Wick et al. [37]. The rate of the mean value of two target cells per day is given by $\frac{2.4 \text{ cells}}{24 \text{ h}}$ = 0.1 cells/h. At the time of experiment, they have taken target cells $5 \times 10^5$ cells/ml in 2 ml. wells. We have already mentioned that the value of $k_1$ have been taken from [26], which we need for the calculation of $\alpha_2$ (killing rate of CD8+ T cells). The value of $k_1$ to be taken as $2.7 \times 10^4$ cells/ml and multiplied by the volume of the well and then substituting the values into $\alpha_2 \frac{G}{G+k_1}$ = 0.09167 h$^{-1}$ $\Rightarrow \alpha_2$ = 0.11642 h$^{-1}$ $\approx 0.12$ h$^{-1}$.

We have estimated the values of the growth rate ($r_2$) of macrophages, the activation rate ($a_1$) of macrophages due to interaction with the IFN-$\gamma$ and elimination term ($\alpha_3$) of the macrophages due to interaction with glioma cells are 0.3307 h$^{-1}$, 0.1163 h$^{-1}$ cells.pg and 0.0194 h$^{-1}$ respectively. We took the value of $M_{max}$ (the carrying capacity of macrophages) is $10^6$ cells from the experimental value of Gutierrez et al. [38].

**Table 1. Parameter values used for numerical simulations.**

| Parameters | Description | Values | Units | Source |
|---|---|---|---|---|
| $r_1$ | growth rate of glioma (tumor) cells | 0.01 | $h^{-1}$ | Estimated |
| $G_{max}$ | carrying capacity of glioma (tumor) | $8.8265 \times 10^5$ | cell | Estimated |
| $e_1$ | Michaelis Menten constant | $10^4$ | pg | [32] |
| $\alpha_1$ | kill rate of macrophage | 1.5 | $pg.h^{-1}$ | Estimated |
| $\alpha_2$ | kill rate of CD8+ T cells | 0.12 | $pg.h^{-1}$ | Estimated |
| $k_1$ | half saturation constant | $2.7 \times 10^4$ | cell | [26] |
| $r_2$ | growth rate of Macrophages | 0.3307 | $h^{-1}$ | Estimated |
| $M_{max}$ | carrying capacity of Macrophages | $10^6$ | cell | [38] |
| $a_1$ | activation rate of Macrophages | 0.1163 | $cell.h^{-1}$ | Estimated |
| $k_4$ | half saturation constant | $1.05 \times 10^4$ | pg | [39] |
| $e_2$ | Michaelis Menten constant | $10^4$ | pg | [32] |
| $\alpha_3$ | death rate of Macrophage | 0.0194 | $h^{-1}$ | Estimated |
| $k_2$ | half saturation constant | $2.7 \times 10^4$ | cell | [26] |
| $a_2$ | antigenicity of glioma (tumor) | $0 - 0.5$ | $h.pg^{-1}$ | [35] |
| $k_5$ | inhibitory parameter | $2 \times 10^3$ | pg | Estimated |
| $\mu_1$ | natural death of CD8+ T cells | 0.0074 | $h^{-1}$ | Estimated |
| $\alpha_4$ | death rate of CD8+ T Cells | 0.1694 | $h^{-1}$ | Estimated |
| $k_3$ | half saturation constant | $3.34452 \times 10^5$ | cell | Estimated |
| $s_1$ | constant source of TGF-$\beta$ | $6.3305 \times 10^4$ | $pg.h^{-1}$ | Estimated |
| $b_1$ | release rate per glioma cell | $5.70 \times 10^{-6}$ | $pg.cell^{-1}.h^{-1}$ | Estimated |
| $\mu_2$ | natural death of TGF-$\beta$ | 6.93 | $h^{-1}$ | Estimated |
| $b_2$ | release rate per CD8+T cell | $1.02 \times 10^{-4}$ | $pg.cell^{-1}.h^{-1}$ | Estimated |
| $\mu_3$ | degradation of IFN-$\gamma$ | 0.102 | $h^{-1}$ | Estimated |

doi:10.1371/journal.pone.0123611.t001

The half saturation of IFN-$\gamma$ was obtained from the half saturation of IFN-$\gamma$ during the activation of macrophages [39], which multiplied by the volume of Central Nervous System, 150 ml. Therefore, $k_4 = 70$ pg/ml $\times$ 150 ml = 10500 pg = $1.05 \times 10^4$ pg. Also, we took the value of $e_2$ (the dependence of CD8+T cell efficiency on TGF-$\beta$) to the order of magnitude of the base line 60.9 pg/ml. from [32], which multiplied by the volume of CNS 150 ml. Therefore, $e_2 = 60.9$ pg/ml $\times$ 150 ml = 9135 pg $\approx 10^4$ pg. By the experiment on a separate culture disc by Mukherjee et al. [26], we took the value of $k_2$ (half saturation constant) as $2.7 \times 10^4$ cells.

All the parameter values are put in tabular form (see Table 1).

## Results

The model, given by (Eq 6), is subjected to numerical simulation for retrieving potential therapeutic effects using T11TS. The aim of the numerical simulation is to show what this model would predict under different experimental scenarios, similar to reported in Mukherjee et al. [26]. These simulations explain some of the reported experimental observations.

### Before T11TS Administration

Figs (4–6) show the dynamics of the three state variables, namely, malignant glioma cells, macrophages and CD8+ T-cells before the administration of T11TS. From the figures, it is clear that the body's own defence mechanism fails to control the growth of malignant glioma cells. As observed in Fig 4A, there is a steep growth of malignant glioma cells for 180 days (6

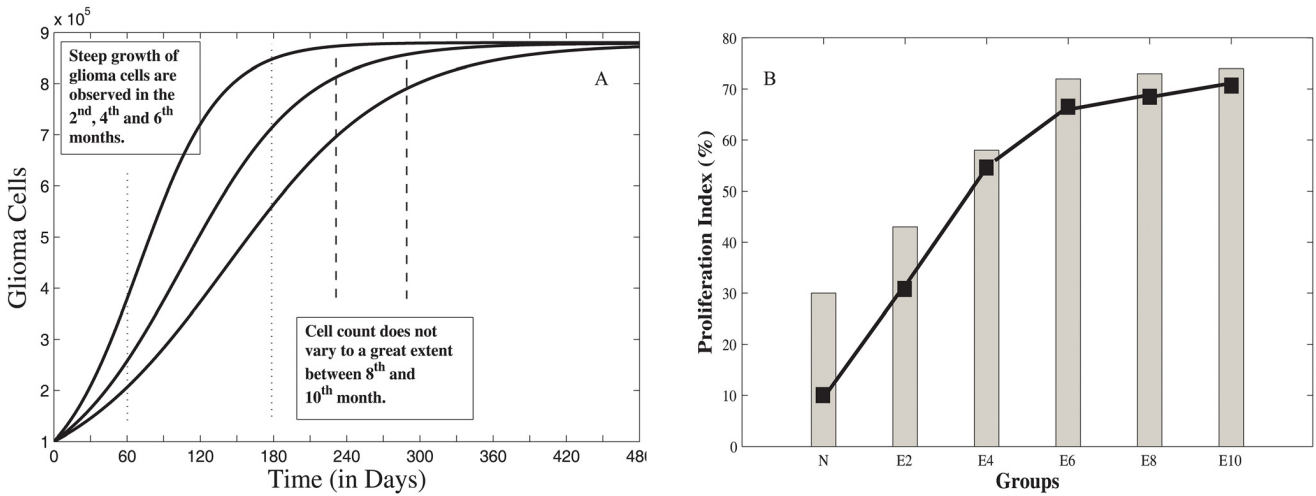

**Fig 4. Glioma cell dynamics without T11 target structure (T11TS).** The left panel A shows the growth of malignant glioma from model simulation and the right panel B gives the experimental data showing glioma cell proliferation index (N to E10).

months) and then the proliferation of glioma cells becomes slow and achieves a state of saturation. The cell count did not vary to a great extent between (240–300) days (between 8th and 10th months), which is in good agreement, when compared with the tumor cell proliferation index of ENU injected animal brain cells (see Fig 4B, namely, E2, E4, E6, E8, E10), counted on 2nd, 4th, 6th, 8th and 10th months respectively. The progressive increase of glioma cell population indicates initiation of uncontrolled cell division. The process of gradual changes from normal to neoplastic stage and the ultimate establishment of intracranial neoplasm occurs during this period, which implies that the probability of the survival rates of the rats decreases during this period [26].

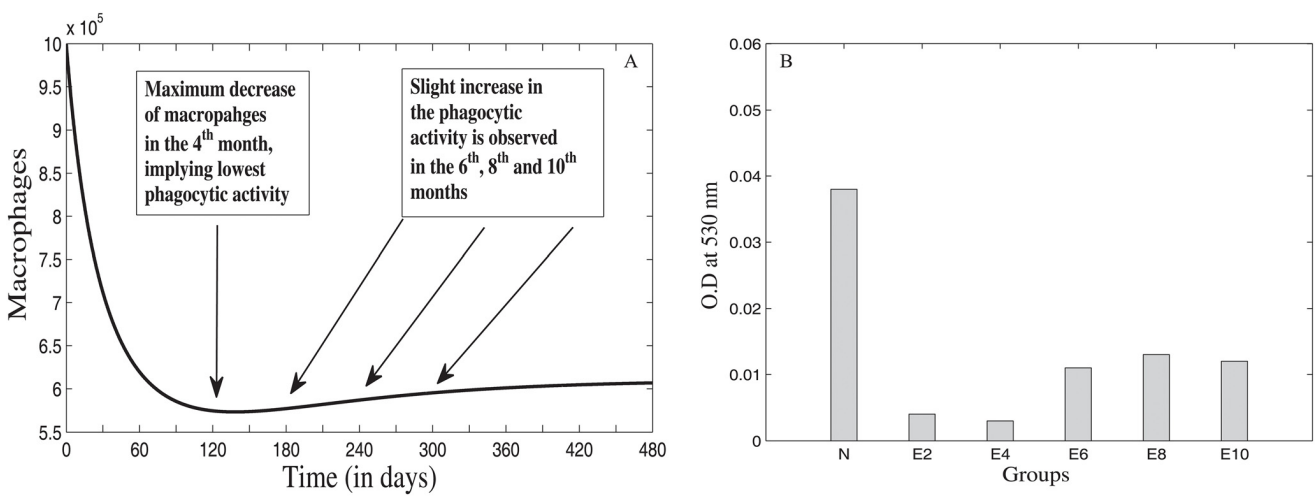

**Fig 5. Dynamics of Macrophages before T11 target structure (T11TS) administration.** The left panel A gives the model simulation showing the decrease in the cell count of macrophages and hence in its phagocytic activity. The right panel B gives the experimental data showing phagocytic activity of macrophages (N to E10).

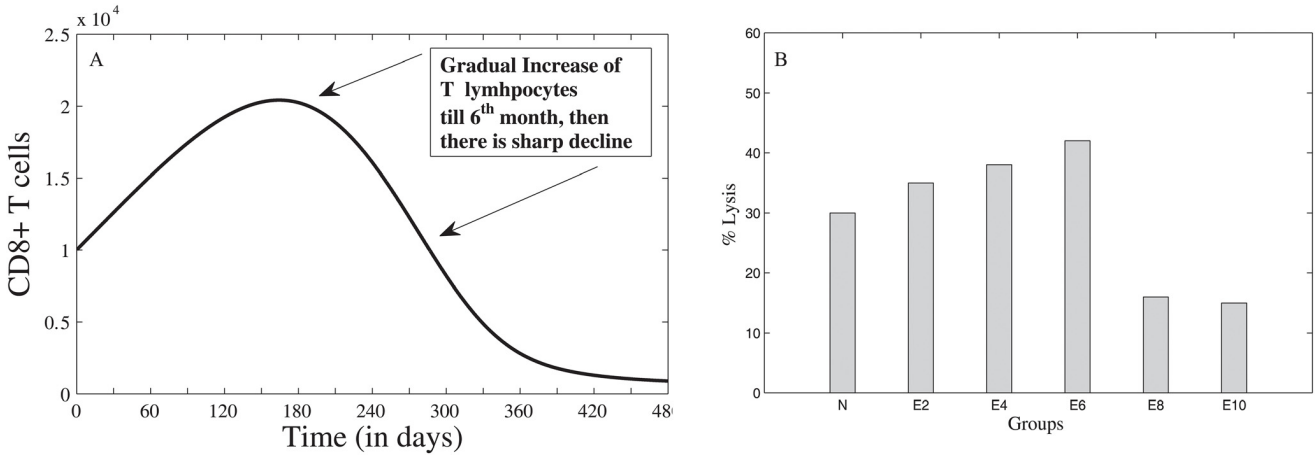

**Fig 6. Dynamics of CD8+ T cells before administration of T11TS.** The left panel A gives the model simulation showing the attempt and failure of CD8+ T cells to counter-attack the malignant glioma cells. The right panel B gives the experimental data showing the decrease in cytotoxic efficacy (N to E10) of CD8 + T cells.

Fig 5A shows the dynamics of macrophage behavior before the administration of T11TS. From the figure, we observe that the cell count of macrophages decreases from its initial value to a significant level in two months (60 days); then a gradual decrease follows till the lowest value is reached in the 4th month (120 days); after that there is a slight increase in the number of macrophages in the 6th (180 days), 8th (240 days) and 10th (300 days) months respectively and then, it reaches a steady state value. This means that the phagocytic activity of the macrophages reduces drastically in the first two months of tumor progression without T11TS, the maximum decrease occurring in the 4th month (see Fig 5B), which may be due to the decrease in the reactive oxygen production [26]. But, there is a slight increase in the macrophage mediated phagocytosis in the 6th, 8th and 10th months, which may be interpreted as a macrophage resistance against the malignant glioma cells.

Fig 6A shows the dynamics of CD8+ T cells and it is observed from the figure that the cell count increases gradually till 180 days (till the 6th month) and then, there is a sharp decline, which points to the fact that though T-lymphocytes tries to counterattack the glioma cells, its cytotoxic efficacy sharply decreases in the 8th month with maximum decrease observed in the 10th month (see Fig 6B). This may be due to the action of the inhibitory cytokine feedback mechanism (say, TGF-$\beta$) exerted by the fully grown tumor mass [40].

## After T11TS Administration

Fig 7 shows the dynamics of malignant glioma cells after the administration of T11TS. The first dose of T11TS was injected after 7 months on ENU injected animals, which shows a significant decrease in the cell count of malignant glioma cells. This agrees with the cell proliferation index of glioma cells obtained from experiment (see Fig 7D for E10 to ET1). This decrease may be interpreted as the increase of total survival period. A sharp increase of glioma cells is noticed after 250 days. Here, we predict that if the 2nd dose of T11TS is not administered soon, tumor resurgence will occur, which tries to reach a steady state as seen before T11TS administration and this is evident from Fig 7A. The second dose, which is administered after 6 days, does not exhibit such a profound effect but still there is a decline in the cell count of glioma cells (Fig 7B). After another 6 days, when the third dose of T11TS is administered, the cell count goes to

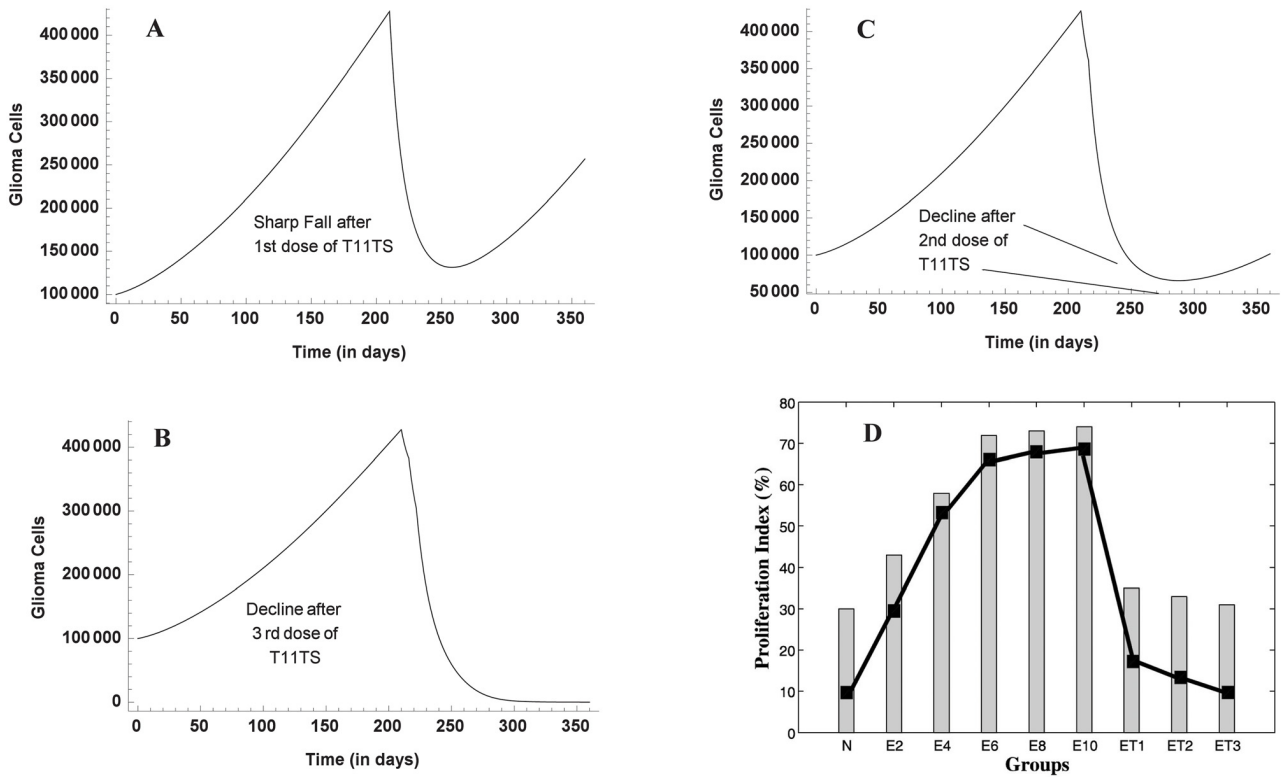

**Fig 7. Malignant glioma dynamics after the doses of T11 target structure.** Through model simulation, the three figures (A,B,C) show the behavior of malignant glioma cells after the (A) first dose (B) second dose (C) third dose of T11TS, the first being administered in the 7th month (210th day), followed by the other two at an interval of 6 days. The fourth figure (D) shows the experimental data of proliferation index of glioma cells after T11TS administration (N to ET3).

doi:10.1371/journal.pone.0123611.g007

zero, thereby succeeding in eradicating the malignant glioma cells (Fig 7C). The simulation agrees with the experimental data as shown in Fig 7D (ET1, ET2, ET3).

Enhanced phagocytic activity occurs after the first dose of T11TS (administered on 210th day) hinting at the probable arming of the macrophages against the malignant glioma cells. The cell count of macrophages increases significantly, thereby facilitating high macrophage stimulation for malignant glioma cell phagocytosis. However, the 2nd and 3rd doses (administered on the 216th and 222nd day respectively) do not have significant stimulation when compared to the first dose (Fig 8). This dynamics of macrophages after the doses of immunostimulatory agent T11TS suggests lymphokine mediated activation of macrophages may occur in a dose dependent manner, which agrees with the experimental data obtained from [26] (see inset Fig 8 for ET1, ET2, ET3).

The cytotoxic efficacy of CD8+ T cells take an upper hand after the 1st and 2nd dose of T11TS, indicating significant lymphocyte proliferation and activation but most significant increase in the cell count of CD8+ T cells and hence improvement in the cytotoxic activity of the lymphocytes (CD8+ T cells), is observed after the administration of the 3rd dose of T11TS (Fig 9), which actually helps in eliminating malignant glioma cells. This is in good agreement with the experimental data (see inset Fig 9 for ET1, ET2, ET3) obtained from [26].

TGF-$\beta$, the immunosuppressive factor secreted by the brain tumor increases sharply between (0–180) days (Fig 10A), which suppresses the activation and proliferation of

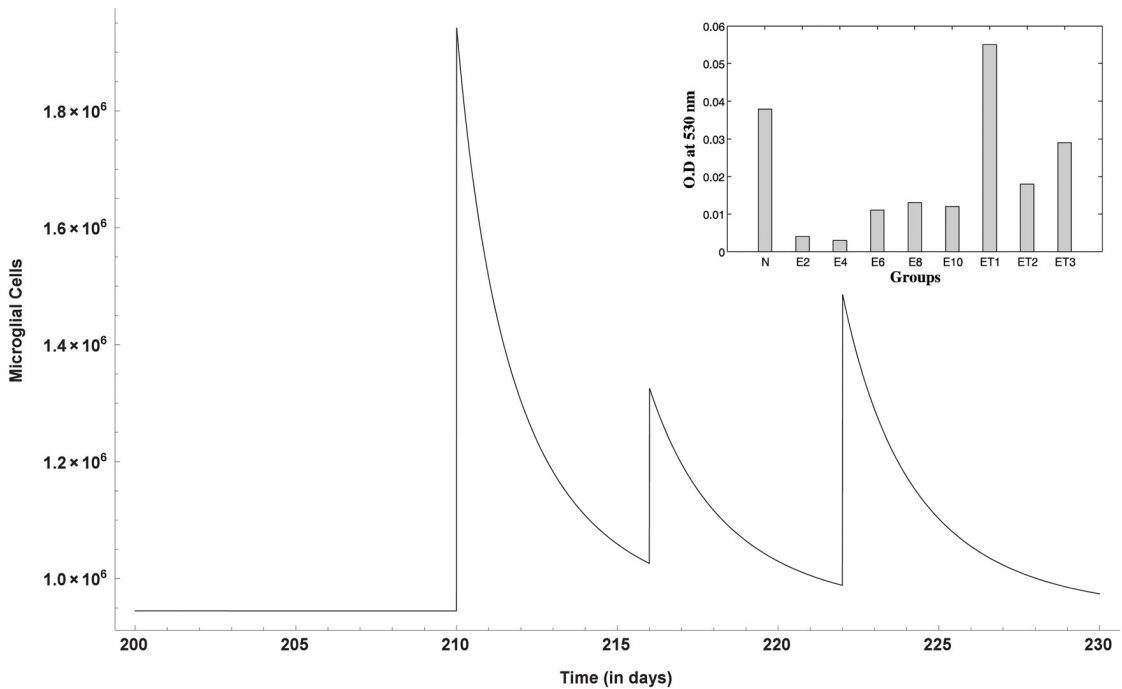

**Fig 8. Macrophage stimulation increases with the doses of T11TS.** Model simulation shows significant stimulation of macrophages after the first dose, when compared with the next two doses. The inset figure represents the experimental data showing increase in phagocytic activity of macrophages.

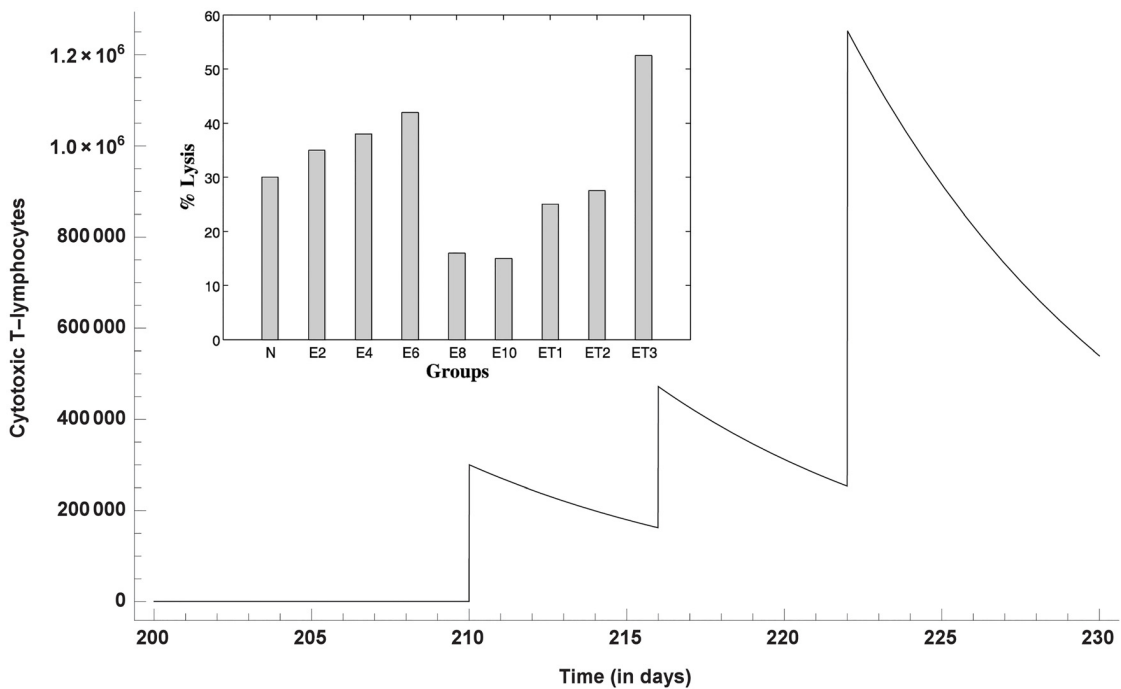

**Fig 9. Dynamics of CD8+ T-cells after the administration of T11 target structure.** Optimal increase in the cell count and therefore efficacy of CD8+ T cells is observed after the 3$^{rd}$ dose of T11TS. The inset figure shows the experimental data of cytotoxic efficacy.

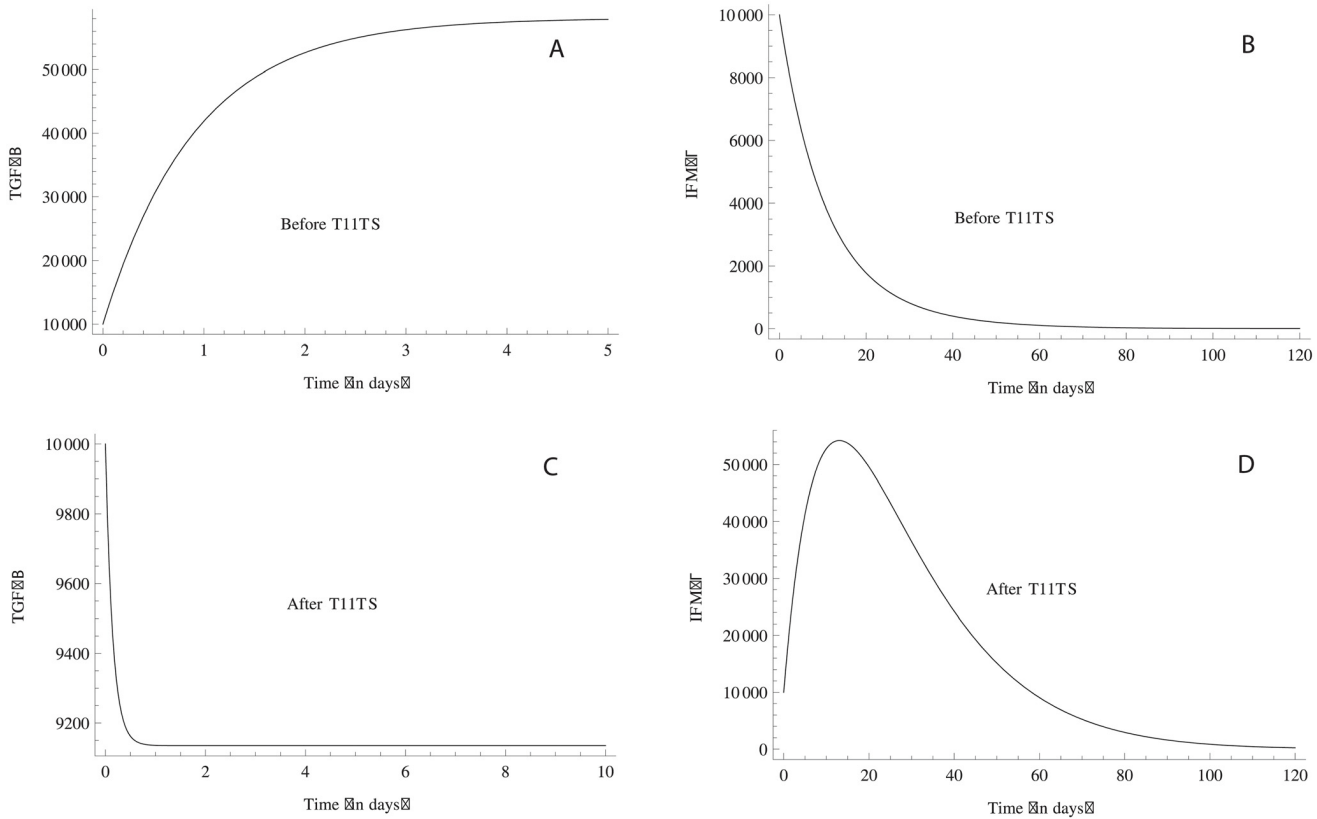

**Fig 10. Dynamics of TGF–β and IFN–γ.** Before the administration of T11TS, TGF–β, the immunosuppressive agent, suppresses the activation and proliferation of immune components with its increase (Fig. A) whereas IFN–γ degrades in 2 months, pointing out its failure to activate the immune system (Fig. B). However, after the doses of T11TS, TGF–β decreases and immunostimulatory effect of IFN–γ increases (Fig. C and Fig. D).

doi:10.1371/journal.pone.0123611.g010

macrophages and CD8+ T cells, resulting in the steep growth of the glioma cells in the 2<sup>nd</sup>, 4<sup>th</sup> and 6<sup>th</sup> months (0–180 days) (see Fig 4A). But, this down-regulation is somewhat balanced by IFN-γ, which increase T-lymphocyte migration across blood-brain-barrier (BBB) [41] and also upregulates MHC Class II on macrophages. But, without T11TS, IFN-γ fails to revive the immune system to counter-attack the malignant glioma cells and degrades in approximately 2 months (Fig 10B). TGF–β, the immunosuppressive factor decreases significantly after the administration of T11TS and the cytokine IFN-γ is stimulated, which activates the macrophages, featuring its immunostimulatory and immunomudulatory effects (Fig 10C and 10D). Thus, from the above observations, T11 target structure seems to clear the first step in its approach to destroy malignant glioma cells.

## Discussion

In the class of heterogeneous group of cancer, brain tumor or intracranial neoplasm is one of the toughest challenges in the field of medical science. Despite multi-modal therapeutic efforts, patient's mean survival for Grade-IV tumor is only 9–12 months and 2 years for Grade-III tumors [3]. Due to its unique anatomy and biology, conventional treatment strategies fails in malignant brain tumors (gliomas). Therefore, adaptation of different immunotherapeutic strategies are required as an alternative modality of glioma treatment. T11 target structure, a transmembrane glycopeptide of sheep red blood cells, when administered in experimentally induced brain tumor (glioma) animals, was found to exert a profound immunopotentiatory effect on peripheral immunocytes including lymphocytes and macrophages. One of the major advantage of using T11TS as a therapeutic agent is its ability to overcomes BBB impermeability [26, 42]. Motivated by the fact that an efficacious therapeutic method is necessary, we analyze the mentioned result using a mathematical model, whose parameters were estimated from published experimental data.

Though the schematic diagram also shows interactions between brain tumor and the immune system, namely, macrophages, CD8+ T-cells, CD4+ T-cells, TGF-$\beta$, IFN-$\gamma$, microglial cells, dendritic cells etc., we have kept our model simple by choosing only five state variables. The challenging part is to obtain the sensitivity of 23 system parameters and its estimation, which has been done successfully. We know that the growth of glioma cells is a heterogeneous process (which is stochastic in nature) but the details of progressive development (preclinical development, which includes growth pattern and invasive nature) of the malignant glioma cells is not the object of study here. In the present course of investigation, we are interested in the know-how of the primary resistance offered by the immune system after the complete establishment of intracranial neoplasm (after ENU administration) and the subsequent dynamics of the effect of immuno-stimulatory agent T11 target structure.

To examine potential approaches towards achieving glioma eradication with T11 target structure as a potent immune stimulator, we develop a mathematical model involving immune components, namely, macrophages, CD8+ T cells, $TGF - \beta$ and $IFN - \gamma$. The sensitivity analysis has been performed on the system parameters, which indicates $r_1$ (growth rate of malignant glioma cells) to be the most sensitive parameter. The consequent subset selection and parameter identifiability set also contains $r_1$, which has been estimated from pre-clinical rat experiment of the growth of malignant glioma cells in absence of T11 target structure. Thus, if no treatment is administered, the model predicts that the malignant glioma grows to a large size, either at system's carrying capacity or close to it. Inevitably, the untreated malignant brain tumor will grow to its maximum size, since the natural immune response against it, is too weak. After T11TS administration, the cytotoxic activity of CD8+ T-lymphocytes is greatly enhanced, indicating potentiation of intracranial immune responses. The macrophages also get activated and there is an increase in the cell count, which felicitates the phagocytic activity of macrophages so that the malignant glioma cells are eradicated. An important advantage of local intro-tumoral infusion of T11TS is their mobility, which can penetrate through areas of brain. This lowers the level of tumor produced TGF-$\beta$, which, in turn, improve the host's immune response to the malignant glioma cells.

The main conclusion of the present study is that our model emulates some of the the experimental findings of Mukherjee et al. [26]. From the results obtained, we can say that the interaction of T11 target structure plays an important role in T-cell and macrophage activation and proliferation. While examining the dynamics of the model, two biologically important observation emerged. First, with the administration of T11 target structure, the glioma cell proliferation is controlled, which increases the survival rate of the animals. Second, with the increase in macrophages and CD8+ T cell count after T11TS administration, our model predicts the increase in phagocytic activity of macrophages as well as cytotoxic efficacy of the CD8+ T cells, which is in good agreement with Mukherjee et al. [26]. The proposed model has the potential to predict and quantify the doses of T11 target structure in other bigger animals (say, chimpanzee). Several other factors or variables, say microglia (a strong candidate for the clearance of cells undergoing apoptosis) or dentric cell or CD4+ T-cells may be introduced in the model and the corresponding dynamics may be studied, which may later be verified with the experimental data. We sincerely hope that immunotherapy with T11 target structure turns out to be a promising therapeutic method and needs to be investigated on humans.

## Acknowledgments

Sandip Banerjee thanks Prof. Hien Tran (North Carolina State University, Raleigh, USA) for useful discussion on sensitivity analysis and parameter estimation. Subhas Khajanchi acknowledges support from the Ministry of Human Resource Development (MHRD), Government of India.

## Author Contributions

Conceived and designed the experiments: SC. Performed the experiments: SC. Analyzed the data: SC SB SK. Contributed reagents/materials/analysis tools: SC SK SB. Wrote the paper: SK SB SC.